\documentstyle[prl,aps,epsf,twocolumn]{revtex}  % DON'T CHANGE
\begin{document}                % INITIALIZE - DONT CHANGe
\title{Long-time dynamics of spontaneous parametric down-conversion
and quantum limitations of conversion efficiency}
\author{Michael Fleischhauer and Oliver Veits}
\address{Sektion Physik, Ludwig-Maximilians-Universit\"at M\"unchen, 
Theresienstra\ss e 37, D-80333 M\"unchen, Germany}
% address. (Remove the left % marks)
%
\date{\today}

\maketitle

\tighten

\begin{abstract}                % DON'T CHANGE THIS LINE
We analyze the long-time quantum 
dynamics of degenerate parametric down-conversion
from an initial sub-harmonic vacuum (spontaenous down-conversion).
Standard linearization of the Heisenberg equations of motions fails
in this case, 
since it is based on an expansion around
an unstable classical solution and neglects pump depletion.
Introducing a mean-field approximation we find
a periodic exchange of energy between the pump and subharmonic mode 
goverened by an anharmonic pendulum equation. From this equation the
optimum interaction time or crystal length for 
maximum conversion can be determined.
A numerical integration of the 2-mode Schr\"odinger 
equation using a dynamically optimized basis of displaced and 
squeezed number states verifies the characteristic times predicted by the
mean-field approximation. In contrast to semiclassical and  mean-field
predictions it is found that  
quantum fluctuations of the pump mode lead to a substantial limitation
of the efficiency of parametric down-conversion.
\end{abstract}


\narrowtext

%%%%%%%%%%%%%%%%%%%%%%%%%%%%%%%%%%%%%%%%%%%%%%%%%%%%%%%%%%%%%%%%%%%%%%%%%

\section{introduction} 

%%%%%%%%%%%%%%%%%%%%%%%%%%%%%%%%%%%%%%%%%%%%%%%%%%%%%%%%%%%%%%%%%%%%%%%%%

Owing to its relative simplicity but yet richness, the 
process of parametric down-conversion is one of the most intensively studied
in quantum optics \cite{Louisell61,Mollow67,Tucker69,Milburn81,Hillery84}.
Here photons of a coherent pump field are transformed into pairs of 
signal and idler photons \cite{Bloembergen,Shen}
which can display nonclassical quantum correlations \cite{Heidmann87}
or perfect squeezing in the case of degeneracy. We here restrict ourselves 
to the latter situation, where both down-converted photons are emitted 
into the same radiation mode. A standard approach to analyze the
quantum fluctuation in nonlinear optical system is to assume small fluctuations
around the classical solutions, i.e. to  linearize the Heisenberg 
equations of motion . The linearization approximation fails however in the case
of a vacuum input of the sub-harmonic mode, since it
neglects pump depletion and is thus  
only valid for an infinite input intensity of the pump field. 
Thus  linearization can neither be used to study the effect of 
finite system size, i.e.
finite pump intensity nor the long-time dynamics of the parametric process. 

Using a short-time perturbation expansion, Crouch and Braunstein
analyzed the leading order corrections to the maximum degree of squeezing
due to finite pump intensities \cite{Crouch88}.
Here we are interested in the long-time behaviour of parametric down-conversion.
In particular we aim to determine the optimum interaction time 
(propagation length in the crystal) for maximum down-conversion 
and the maximum efficiency of this process.
In the case of a vacuum input of the sub-harmonic mode,
both quantities are goverened by quantum effects. 
We find that in contrast to the classical predictions, these 
quantum effects limit the 
maximum conversion efficiency from a pump photon into two sub-harmonic 
photons to a value much less than unity.   
This limitation could be of importance for applications in
quantum communication and cryptography on the single photon level.

%%%%%%%%%%%%%%%%%%%%%%%%%%%%%%%%%%%%%%%%%%%%%%%%%%%%%%%%%%%%%%%%%%%%%%%%%

\section{model, classical dynamics and linearisation}

%%%%%%%%%%%%%%%%%%%%%%%%%%%%%%%%%%%%%%%%%%%%%%%%%%%%%%%%%%%%%%%%%%%%%%%%%

In order to describe stationary parametric conversion of travelling-wave 
pump radiation into travelling-wave sub-harmonic radiation we introduce a
moving coordinate system. Ignoring transversal 
degrees of freedom we find the following Heisenberg equations
of motion
\begin{eqnarray}
\frac{\rm d}{{\rm d} t} a_1 &=& K\, a_2 a_1^\dagger,\label{a1}\\
\frac{\rm d}{{\rm d} t} a_2 &=& - \frac{K^*}{2} a_1^2.\label{a2}
\end{eqnarray}
$a_1$ and $a_2$ are the bosonic mode operators of the sub-hamonic
and pump fields respectively. 
$K$ describes the strength of the nonlinear process. 
It is proportional to the nonlinear susceptibility $\chi^{(3)}$
of the crystal and the inverse of the beam diameter. The  time 
evolution in the
moving frame corresponds to a spatial evolution in the lab frame
and the fields at $t=0$
are the input fields. 
Due to the phase symmetry of the equations
\begin{eqnarray}
a_1 &\to& \pm a_1\, {\rm e}^{i\phi_1},\nonumber\\
a_2 &\to& \enspace a_2\, {\rm e}^{i\phi_2},\nonumber\\
K   &\to& \enspace K\, {\rm e}^{i(2\phi_1-\phi_2)}\nonumber
\end{eqnarray}
we may choose $K$ and the initial amplitude of the pump field
$\langle a_2(t=0)\rangle$ real. 
The equations of motion (\ref{a1},\ref{a2}) obey the Manley-Rowe
relation \cite{Bloembergen,Shen}, which 
states that the total energy of the free (!) system
is conserved.
\begin{equation}
\frac{\rm d}{{\rm d} t} \langle a_1^\dagger a_1\rangle +
2 \frac{\rm d}{{\rm d} t} \langle a_2^\dagger a_2\rangle =0.\label{ManleyRowe}
\end{equation}

Even though Eqs.(\ref{a1}) and (\ref{a2}) seem simple, the nonlinearity
prevents an analytic solution of the quantum problem. Therefore approximations
are necessary. A frequently used approximation is the linearisation around
the classical solutions. In order to discuss the validity of 
this approximation, let us first
consider the classical problem, where the Bose operators $a_1$ and $a_2$
are replaced by c-numbers $\alpha_1$ and $\alpha_2$. 
\begin{eqnarray}
\frac{\rm d}{{\rm d} t} \alpha_1 &=& K\, \alpha_2 \alpha_1^*,\label{alpha1}\\
\frac{\rm d}{{\rm d} t} \alpha_2 &=& - \frac{K}{2} \alpha_1^2.\label{alpha2}
\end{eqnarray}
One clearly sees, that for vanishing sub-harmonic input, i.e. $\alpha_1(0)=0$,
both amplitudes remain constant. This solution is linearly unstable
and any fluctuation will be exponentially amplified. 
The time evolution critically 
depends on the amplitude and phase of an initial classical fluctuation. 
Thus a classical calculation cannot determine the optimum interaction
time (or crystal length) for maximum conversion. 

In the standard linearization approach, the pump-mode operator
is replaced by its classical input amplitude. This turns the quantum 
problem into a linear one, which can immediately be solved. One finds that
the time-evolution operator of the sub-harmonic mode is given by 
\begin{equation}
{\sf U}_{\rm lin}(t)={\sf S}[\eta(t)],
\end{equation}
where {\sf S} is the so-called squeezing operator \cite{Yuen76}
\begin{equation}
{\sf S}[\eta]=\exp\Bigl\{\frac{\eta}{2} a_1^{\dagger 2} 
-\frac{\eta^*}{2} a_1^2\Bigr\}
\end{equation}
with a squeezing parameter that grows linear with time
\begin{equation}
\eta(t) = K\alpha_2 t.
\end{equation}
Since $K$ and $\alpha_2$ have been choosen real, the time evolution
will lead to a squeezing of the fluctuations of the out-of-phase
component of the sub-harmonic mode $p_1$, ($a_1=x_1+i p_1$) 
below the standard vacuum limit.
The quantum noise of $p_1$ monotonously decreases with time and 
simultaneously the quantum noise of $x_1$ increases. 
The increase of the fluctuations in the in-phase component $x_1$ is
associated with a steady increase of the sub-harmonic photon number
\begin{equation}
\langle a_1^\dagger a_1\rangle = \sinh^2 \eta(t) =\sinh^2 K\alpha_2 t.
\end{equation}
This result violates the Manley Rowe relations (\ref{ManleyRowe})
and  indicates the breakdown of the linearization for larger
times. The growing fluctuations of the (anti-squeezed component of the)
sub-harmonic mode can at some point not assumed to be small anymore. 
They will lead to a decrease (depletion) of the pump-mode
amplitude and to fluctuations in this mode.

%%%%%%%%%%%%%%%%%%%%%%%%%%%%%%%%%%%%%%%%%%%%%%%%%%%%%%%%%%%%%%%%%%%%%%%%

\section{mean-field approximation and optimum interaction time}

%%%%%%%%%%%%%%%%%%%%%%%%%%%%%%%%%%%%%%%%%%%%%%%%%%%%%%%%%%%%%%%%%%%%%%%%

As noted above a linearization of the Heisenberg equations of motion
cannot be used to study the long-time behaviour of spontaneous 
(vacuum input) parametric down-conversion. The quantum fluctuations
of the sub-harmonic mode and their backaction onto the pump mode
are essential and need to be taken into account.
We may however replace
the pump-mode amplitude by its average value, which amounts to
a mean-field approximation \cite{Veits98}. 
With this we obtain the equations of motion
\begin{eqnarray}
\frac{\rm d}{{\rm d} t} a_1 &=& K\, \langle a_2\rangle a_1^\dagger,
\label{mf_a1}\\
\frac{\rm d}{{\rm d} t} \langle a_2\rangle  &=& 
- \frac{K^*}{2} \langle a_1^2\rangle.\label{mf_a2}
\end{eqnarray}
Thus we have transformed the original set of nonlinear operator
equations into a {\it linear} operator equation plus a nonlinear
classical one. One easily verifies that  equations
(\ref{mf_a1}) and (\ref{mf_a2}) obey the Manley-Rowe relation.
\begin{equation}
\frac{\rm d}{{\rm d}t} \langle a_1^\dagger a_1\rangle =
K\langle a_2^\dagger\rangle \langle a_1^2\rangle + c.c. = 
- 2\frac{\rm d}{{\rm d} t} \langle a_2^\dagger a_2\rangle.
\end{equation}
The mean-field equations correpond to a time-evolution operator
\begin{equation}
{\sf U}_{\rm mf}(t)=
{\sf D}_2\bigl[\beta(t)\bigr]\, {\sf S}_1\bigl[\eta(t)\bigr],
\label{U}
\end{equation}
that consists of a coherent displacement operator for the
pump mode and a squeezing operator for the sub-harmonic mode.
\begin{eqnarray}
{\sf D}(\alpha)& =& \exp\bigl\{\alpha a^\dagger - \alpha^* a\bigr\},\nonumber\\
{\sf S}(\eta) &=& \exp\Bigl\{\frac{\eta}{2} a^{\dagger 2} 
-\frac{\eta^*}{2} a^2\Bigr\}.
\nonumber
\end{eqnarray}
Thus the interaction leads to a shift of the 
coherent amplitude of the pump mode by the amount
\begin{equation}
\beta(t)= -\frac{1}{2} K \int_0^t\!\!\! {\rm d}t^\prime\, 
\langle a_1^2(t^\prime)\rangle.\label{beta}
\end{equation}
At the same time the sub-harmonic mode is squeezed
by 
\begin{equation}
\eta(t) = K\int_0^t\!\!\! {\rm d} t^\prime\, \langle a_2(t^\prime)\rangle.
\label{eta}
\end{equation}
In contrast to the linearisation, the squeezing parameter does  not
increase indefinitely, since the pump mode amplitude decreases, characterized
by the displacement parameter $\beta$.
 $\beta(t)$ and $\eta(t)$ are not independent. From the
mean-field equations we find
\begin{equation}
\ddot\eta(t) = K \dot\beta(t).\label{beta2}
\end{equation}

If we know $\eta(t)$ we can immediately obtain the amplitude
of the (classical) pump mode from Eq.(\ref{eta}). On the other hand
we find the following coupled equations for the sub-harmonic photon
number and correlation function
\begin{eqnarray}
\frac{\rm d}{{\rm d}t} \langle a_1^\dagger a_1\rangle &=& 
2 \dot\eta\, \langle a_1 a_1\rangle,\nonumber\\
\frac{\rm d}{{\rm d}t} \langle a_1 a_1\rangle &=& 
\dot\eta\,\Bigl(2 \langle a_1^\dagger a_1\rangle +1 \Bigr),\nonumber
\end{eqnarray}
which have the solutions
\begin{eqnarray}
\langle a_1^\dagger a_1\rangle &=& \sinh^2\eta,\label{sol1}\\
\langle a_1 a_1\rangle &=&\frac{1}{2} \sinh 2\eta.\label{sol2}
\end{eqnarray}
Thus the knowledge of $\eta$ is sufficient to determine all relevant 
quantities. 
From (\ref{beta}) and (\ref{beta2}) we find $\ddot\eta = -(K^2/2) 
\langle a_1^2\rangle$ and thus the dynamics of the squeezing parameters
is goverened by an anharmonic pendulum equation.
\begin{equation}
\ddot \eta(t) = -\frac{1}{4} K^2 \sinh 2 \eta(t),\label{pendel}
\end{equation}
with the initial conditions
\begin{eqnarray}
\eta(0) &=& 0,\nonumber\\
\dot\eta(0) &=& K\langle a_2(0)\rangle.\nonumber
\end{eqnarray}
The anharmonic pendulum equation with the given initial
conditions is equivalent to the integrated
Manley-Rowe relation
\begin{equation}
\frac{2}{K^2}\dot\eta^2 + \sinh^2 \eta =  2 n_2^0 = 2|\langle a_2(0)\rangle|^2.
\label{energy}
\end{equation}
This suggests a mechnical analogue.
If $\eta$ is interpreted as the spatial coordinate of a classical particle
moving in one dimension,
the first term in Eq.(\ref{energy}) 
represents its kinetic and the second its potential
energy. In the chosen units the kinetic energy is then 
twice the pump-mode photon number
and the potential energy the photon number of the sub-harmonic mode.

\begin{figure}
\begin{center}
\leavevmode \epsfxsize=7 true cm
\epsffile{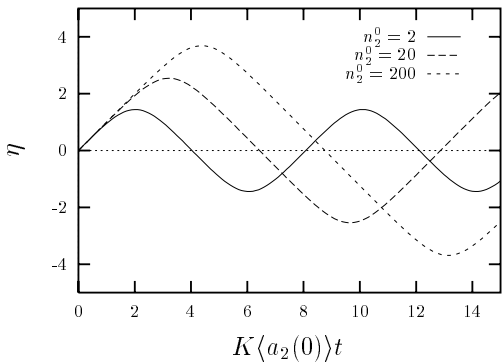}
%\epsffile{squeezingparam.eps}
\end{center}
\caption{Squeezing parameter as function of scaled time for
coherent pump with input intensity $\langle a_2(0)\rangle^2=n_2^0$.
}
\label{sq_parameter}
\end{figure}

Fig.~1 shows the squeezing parameter as function of the scaled time
$K\langle a_2(0)\rangle t$ for different initial photon numbers
$n_2^0:=|\langle a_2(0)\rangle|^2$. 
The squeezing parameter reaches a maximum value and there is
an optimum interaction time or crystal length for maximum squeezing.
The corresponding optimum time is a quarter of the oscillation
period in the anharmonic potential.
\begin{equation}  
KT_{\rm sq}= \frac{1}{2}\int_0^{y_{\rm max}} \!\!\! \frac{ {\rm d} y}
{\sqrt{n_2^0- \frac{1}{2} \sinh^2 y}}\approx \frac{1}{4}\,{\rm ln}\, \Bigl[
 n_2^0\Bigr]
,\label{sq_time}
\end{equation}
where $\sinh^2 y_{\rm max} = 2 n_2^0$. 
This results agrees with that of the short-time perturbation
expansion by Crouch and Braunstein \cite{Crouch88}. A comparision with the 
Crouch-Braunstein result shows however that the maximum 
amount of noise
reduction found in  mean-field approximation
\begin{equation}
\langle \Delta p_1^2\rangle_{\rm min} = \frac{1}{32 n_2^0}
\end{equation}
is too small. The mean-field approach neglects the fluctuations of the
pump mode, in particular its phase noise. When this is taken into account
the minimum fluctuations are only $\langle \Delta p_1^2\rangle_{\rm min} =
1 / 8 \sqrt{n_2^0}$ \cite{Crouch88,Kinsler93}. 

Maximum conversion of pump into sub-harmonic photons 
is achieved when $\dot\eta=0$, i.e. at the turning points
of the classical pendulum motion. Thus the optimum conversion time 
$T_{\rm conv}$ or equivalently the optimum crystal length
is determined by 
\begin{equation}
K\, T_{\rm conv} = \int_0^{y_{\rm max}} \!\!\! \frac{ {\rm d} y}
{\sqrt{n_2^0- \frac{1}{2} \sinh^2 y}}\approx  \frac{1}{2}\, {\rm ln} \,
\Bigl[n_2^0\Bigr]
,\label{conv_time}
\end{equation}
which is twice the time of maximum squeezing. 
Fig.~2 shows the scaled photon numbers of the pump and sub-harmonic mode
as a function of time.

\begin{figure}
\begin{center}
\leavevmode \epsfxsize=7 true cm
\epsffile{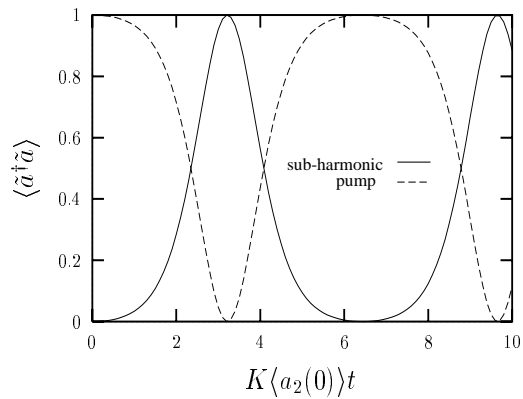}
%\epsffile{photon02.eps}
\end{center}
\caption{Scaled mean photon number of pump (dashed) and sub-harmonic
mode (line) as function of scaled time for 
$\langle a_2(0)\rangle=\sqrt{20}$}
\end{figure}

Since the mean-field
approach takes into account the energy transfer from the pump mode into
sub-harmonic fluctuations, it correctly
describes the oscillatory energy exchange in parametric 
down-conversion from an initial sub-harmonic vacuum.
This is in contrast to the classical or linearization approximation.
The mean-field approximation also allows to determine the optimum 
interaction time for large squeezing or best 
down-conversion, Eqs.(\ref{sq_time},\ref{conv_time}).   The 
maximum conversion efficiency  is unity.

The underlying assumption of the mean-field approach is a quasi-classical
description of the pump field. This assumption becomes however questionable
at the point of total energy conversion and thus the
maximum conversion efficiency obtained in mean-field
approximation may not be correct. 
To calculate this quantity and to discuss the 
influence of quantum fluctuation in particular of the pump mode
we shall
numerically integrate the two-mode Schr\"odinger equation
in the next section.

%%%%%%%%%%%%%%%%%%%%%%%%%%%%%%%%%%%%%%%%%%%%%%%%%%%%%%%%%%%%%%%%%%%%%%%%

\section{numerical integration of two-mode Schr\"odinger equation
and quantum limit to the conversion efficiency}

%%%%%%%%%%%%%%%%%%%%%%%%%%%%%%%%%%%%%%%%%%%%%%%%%%%%%%%%%%%%%%%%%%%%%%%%

A direct numerical integration of the Schr\"odinger or Liouville 
equation is not a straight forward task for multi-mode problems. 
Unless the interacting
modes contain only very few photons, the standard  Fock-basis expansion
requires the use of a large basis set. For the present problem 
a large basis set is required in both modes since during the interaction
all or almost all photons of the pump mode are converted into
sub-harmonic photons and vice versa. 

To avoid the large-memory requirement of a simple Fock space
expansion one may think of choosing a modified basis adapted to the
problem. For example in the initial phase of the process the pump mode
is in a coherent state $|\alpha_2^0\rangle$. Its photon number
distribution is Poissonian  and thus the required number of
basis states is of the order of $|\alpha_2^0|$, which can be large. 
On the other hand
one can displace the number state basis with the unitary transformation
${\sf D}(\alpha)$  introducing the states
\begin{equation}
|\alpha,n\rangle = {\sf D}(\alpha) |n\rangle\label{coh_basis}
\end{equation}
which form a complete set. Clearly at $t=0$ only a single
state is needed to describe the pump mode if $\alpha=\alpha_2^0$. 
As known from the mean-field approach, the coherent amplitude
of the pump mode decreases during the interaction and the
basis set (\ref{coh_basis}) would soon become ineffective. Thus the
parameter $\alpha$ needs to be dynamically adapted, $\alpha\to\alpha(t)$.
 This is easy
to implement in a numerical algorithm that solves the differential
equation. In each time step the expansion coefficients are calculated
in an adapted basis which uses parameters obtained in the previous
time step. These coefficients are then used to update the basis and so on.

If there is no initial symmetry-breaking the coherent amplitude of the
subharmonic mode remains zero at all times. Thus a dynamically adapted
coherent displacement of the sub-harmonic basis states is not useful.
However we have seen in the previous section that the time evolution
of this mode is approximately described by 
a dynamical squeezing ${\sf S}(\eta)$,
see Eq.(\ref{U}). Therefore we expand the state vector of the
sub-harmonic mode in a  squeezed number-basis
\begin{equation}
|\eta,n\rangle = {\sf S}(\eta)\, |n\rangle,
\end{equation}
with a dynamically adapted parameter $\eta=\eta(t)$.

The use of a dynamically optimized squeezed and displaced number
basis \cite{Veits98} 
allowed a numerical integration of the two-mode Schr\"odinger
equation for input photon numbers up to several thousands. 
In Fig.~3 we have shown the scaled real part of the pump mode
amplitude $(\langle a_2\rangle = x_2 + i p_2)$ and its fluctuations
as a function of the scaled time $\tau=K\langle a_2(0)\rangle t$.
Also shown is the mean-field result. One recognizes good
agreement of the predictions for the optimum conversion time
from both approaches.

\begin{figure}
\begin{center}
\leavevmode \epsfxsize=7 true cm
\epsffile{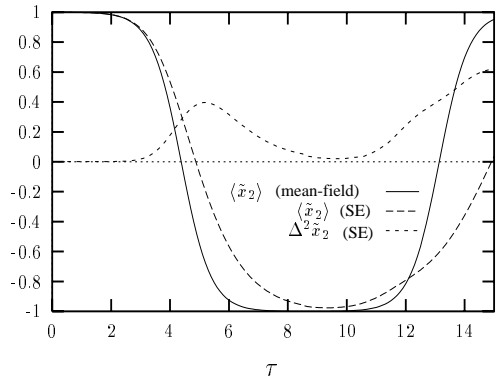}
%\epsffile{x22.eps}
\end{center}
\caption{Scaled in-phase average quadrature component of pump mode 
$\langle \tilde x_2\rangle=\langle x_2\rangle/\langle x_2(0)\rangle$ from
mean-field approximation (line) and numerical integration of two-mode
Schr\"odinger equation (long dashes). Also shown are the fluctuations
of $\tilde x_2$ obtained from numerical integration (short dashes).
$\tau=K\langle a_2(0)\rangle t$, $\langle a_2(0)\rangle =\sqrt{200}$}
\end{figure}

On the other hand, the numerical solution shows,
that at the point of vanishing coherent amplitude of the pump mode,
its fluctuations become large. This implies that the coherent-state
approximation used in the mean-field approach is not valid near the
point of maximum conversion. Furthermore, although the
coherent amplitude vanishes, the mean photon number of the pump remains
finite and thus the conversion efficiency is less than unity. 
Fig.~4 illustrates this.

\begin{figure}
\begin{center}
\leavevmode \epsfxsize=7 true cm
\epsffile{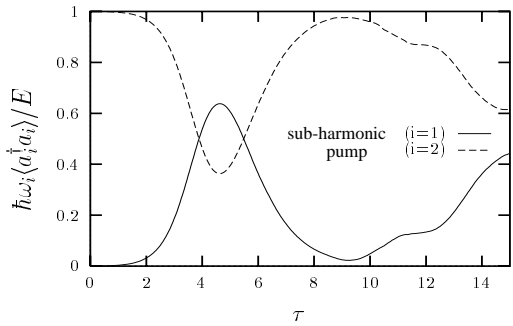}
%\epsffile{n3.eps}
\end{center}
\caption{Time evolution of energy of pump (dashed) and sub-harmonic
mode (line) as function of scaled time $\tau=K\langle a_2(0)\rangle t$
for $\langle a_2(0)\rangle =\sqrt{200}$.}
\end{figure}

Shown are the energies of both fields
as a function of time. One recognizes a maximum conversion of only
about 65\% for 200 input photons of the pump mode. Our calculations indicate
that this value does not increase with increasing input photon number
and is thus not a finite-size effect. Near the point of maximum
conversion the pump-mode amplitude becomes small and the back-action
of the quantum
fluctuations of the sub-harmonic mode (more precisely that of the
anti-squeezed quadrature component) onto the pump mode
gain importance. They lead to an increase of the in-phase quadrature 
fluctuations of the pump field and thus a finite amount of energy
remains in this mode even though the coherent amplitude vanishes.

%%%%%%%%%%%%%%%%%%%%%%%%%%%%%%%%%%%%%%%%%%%%%%%%%%%%%%%%%%%%%%%%%%%%%%%%

\section{summary}

%%%%%%%%%%%%%%%%%%%%%%%%%%%%%%%%%%%%%%%%%%%%%%%%%%%%%%%%%%%%%%%%%%%%%%%%

We have analysed the long-time quantum dynamics of degenerate parametric down
conversion, for which standard approaches like the linearization
of the Heisenberg equations of motion fail. In a mean-field approach,
which assumes a coherent pump mode but takes the sub-harmonic fluctions
fully into account, an oscillatory energy exchanges between the modes is
found. The mean-field approach allows to determine the optimum 
interaction times or crystal lengths for maximum squeezing and
maximum down conversion. Since
this approach neglects the quantum fluctuations of the pump,
it becomes invalid near the point of maximum conversion and cannot
be used to estimate the conversion efficiency. To calculate the
latter we numerically integrated the two-mode Schr\"odinger equation.
The numerical integration was possible for photon numbers 
up to several thousands due to the use of a dynamically optimized,
displaced and squeezed number basis \cite{Veits98}. 
We found that the maximum conversion efficiency is only about 65\%
for a coherent input of the pump mode. This limitation is a pure
quantum effect. The large fluctuations in the anti-squeezed component
of the sub-harmonic field introduce corresponding fluctuations in the
pump mode via the nonlinear interaction. As a result a finite amount
of energy remains in this mode even at the point of vanishing coherent
amplitude.

%%%%%%%%%%%%%%%%%%%%%%%%  begin references %%%%%%%%%%%%%%%%%%%%%%%%%%%%%%

\end{document}